\documentclass[12pt]{iopart}
\begin{document}

\title{Generalized Second Law and phantom Cosmology: accreting black holes}

\author{J.A. de Freitas Pacheco$^1$ and J.E. Horvath$^2$}

\address{$^1$Obsevatoire de la C\^ote d'Azur, Laboratoire Cassiop\'ee,
UMR6202, BP4229, 06304-Nice, France}
\address{$^2$Universidade de S\~ao Paulo, Instituto de Astronomia,
Geof\'\i sica  e Ci\^encias
Atmosf\'ericas, R. do Mat\~ao 1226, Cidade Universit\'aria, S\~ao Paulo, SP, Brasil}
\ead{\mailto{pacheco@oca.eu}, \mailto{foton@astro.iag.usp.br}}

\begin{abstract}

The accretion of phantom fields by black holes within a thermodynamic context
is addressed. For a fluid violating the dominant energy condition, case of a
phantom fluid, the Euler and Gibbs relations permit two different possibilities
for the entropy and temperature: a situation in which the entropy is negative
and the temperature is positive or vice-versa. In the former case, if the
generalized second law (GSL) is valid, then the accretion process is  not allowed whereas
in the latter, there is a critical black hole mass below which the accretion
process occurs. In a universe dominated by a phantom field, the critical mass
drops quite rapidly with the cosmic expansion and black holes are only
slightly affected by accretion. All black holes disappear near the big rip, as suggested
by previous investigations, if the GSL is violated.

\end{abstract}

\maketitle

\section{Introduction}

Data on type Ia supernova distances and other complementary cosmological observations
revealed that presently the expansion of the universe is in an accelerated
phase \cite{riess98} and \cite{perl99}.
The observed acceleration requires the existence of a new component, termed
dark energy, which dominates presently over all other forms of energy and
is characterized by a negative pressure. The dark energy
is frequently modeled as an homogeneous scalar field $\phi$ with a suitable
potential $V(\phi)$. According to the value of the equation of state parameter $w$,
defined by the ratio between the pressure and the energy density ($w=P/\varepsilon$),
three different cases can be distinguished:{\it ``quintessence"}, when $w$ is in the
range $-1 < w < -1/3$ and the kinetic term $\dot\phi^2/2$ is positive;{\it ``cosmological
constant"}, the particular case when $w=-1$ and only the potential term $V(\phi)$
contributes to both the pressure and the energy density of the field; finally, fields
with a negative kinetic term leading to values of $w < -1$, dubbed {\it ``phantom"}
fields. These unusual fields appear in some string \cite{fram03} and super gravity
theories \cite{nil84} and have also some weird properties: they violate the
dominant energy condition ($P+\varepsilon<0$), the energy density increases with
the cosmic time and a universe dominated by a phantom field has a future singularity,
the ``big rip" \cite{cal03}. However, quantum effects may eventually drive the universe
out of the future singularity \cite{nojiri04}.  A further difficulty with
phantom fields concerns quantum instabilities of the vacuum. Processes involving
the graviton-mediated decay of vacuum into two ghost-quanta and two photons
have been studied by \cite{cline04}, who have shown that the divergent nature of the
phase space can only be avoided by imposing a Lorentz
noninvariant momentum cutoff, which cannot guarantee the masslessness of the graviton.
In spite of these theoretical difficulties and of the fact that most cosmological data
be presently in favor of a cosmological constant as the driving expansion acceleration
mechanism, lower values of equation of state parameter, e.g., $w < -1$ cannot
completely be excluded \cite{deu02,cal03,alam06}. In particular the analysis
by \cite{tonry03}, based on supernova
data, assuming a flat universe and adopting a prior based on the Two Degree Field (2dF)
redshift survey constraint on the total matter density parameter, leads to the conclusion
that the equation of state parameter lies in the range {\it $-1.48 < w < -0.72$} at 95\%
confidence level.

Since an accelerated expansion driven by a phantom field remains a possibility,
further consequences related to the presence of such a field in the universe
deserve more detailed investigations, in particular those
related to applications of the generalized second law \cite{bek74} or the entropy bound
\cite{bek81}. A phantom field may completely modify the evolution of black
holes (BHs), since
studies performed by \cite{babi04,babi05} and confirmed later by \cite{gon04} suggest
that all BHs lose mass and disappear completely near the singularity. In this
paper, the effects
of accretion of a phantom fluid by BHs are
revisited in the light of the generalized second law (GSL) and the holographic bound
to the entropy,
both supposed here to hold in a universe dominated by a phantom field.

\section{The evolution of black holes embedded in a phantom field}

\subsection{The generalized second law}

Some studies performed recently on phantom cosmologies and on possible violations either
of the GSL or the entropy bound, are often based on different interpretations
of the GSL, thus
leading to disparate conclusions. Therefore, it seems necessary first to clarify some
aspects related to the entropy and energy conservation in an expanding universe.

In relativistic cosmology, in the absence of entropy sources like
bulk viscosity or particle
production, entropy is a conserved quantity within
a {\it comoving unit volume}. The
total entropy within a volume delimitated by a specific horizon (event, apparent or
causal) is a ill defined quantity. It grows in an open universe but it can decrease
in a closed universe which has already attained the collapsing phase. This does not mean
that the second law of thermodynamics is satisfied in the former case but violated in
the latter, since in both examples the evolution is adiabatic, e.g., no
entropy sources are present inside the considered volume.
Similar considerations can be made for
the energy inside a comoving unit volume.
If the cosmic expansion is adiabatic, then from
the first law
\begin{equation}
\frac{d(\varepsilon a^3)}{dt} + P\frac{da^3}{dt} = 0
\label{firstlaw}
\end{equation}
In a laboratory, an expanding volume loses energy adiabatically to the external world
at the rate given by the eq.~\ref{firstlaw}
and the sum of the energies inside the considered
volume and the outside world (supposed to be delimitated by adiabatic walls)
remains constant.
In an expanding, homogeneous and unbound universe, all comoving regions are
alike in content
and each of them may be regarded as a closed system having no external world
to which the lost energy $-(Pda^3)$ can be transferred, since all regions
experience identical
losses. The usual idea of an expanding volume performing work on its surroundings cannot
apply in this case because the expansion of the space itself is responsible
for the energy losses \cite{har95}.

If BHs are present, they can either accrete energy or emit particles via the Hawking
mechanism, thus producing entropy. In this case, \cite{bek74} conjectured that the
second law holds only for the sum
of the black hole and matter-radiation entropies. This conjecture should
be understood as follows. Consider a thermodynamic system $\Sigma$ consisting of
ordinary matter-radiation
with entropy $S_{m+r}^{init}$ and black holes with entropy $S_{bh}^{init}$, representing
the sum
of the horizon areas inside the system $\Sigma$. The initial total entropy of $\Sigma$ is
\begin{equation}
S_{\Sigma}^{init} = S_{m+r}^{init} + S_{bh}^{init}
\end{equation}
If the initial equilibrium is disrupted because different processes occur as, for instance,
matter-radiation can be accreted by black holes or collapse to form new
black holes, the entropy
of $\Sigma$ will be changed. When the new equilibrium state is established, the total
final entropy of the system will be $S_{\Sigma}^{final}$ and the GSL states that
\begin{equation}
S_{\Sigma}^{final} \geq S_{\Sigma}^{init}
\end{equation}

In phantom cosmologies, since $P+\varepsilon < 0$, the Euler's
relation $Ts = P+\varepsilon$ allows
two alternatives: either the entropy density $s$ is negative, being the
temperature $T$ positive or
the entropy density is positive but in this case the temperature associated to
the phantom fluid
is negative. These two possibilities follow also from the Gibbs relation, assuming that the
energy density is a state function of the temperature and the
integrability condition. In statistical
mechanics, the entropy of a system is a measure of the (logarithm) number of
available states, thus
a positively defined quantity. Other definitions as the Shannon entropy, which
measures the
uncertainty of a discrete random variable in the Information Theory
is also non-negative. However,
since the consequences of both possibilities ($S<0$ and $S>0$) have been examined
in the literature,
here the same procedure will be adopted.
The first possibility leads to the pair of equations
\begin{equation}
s = \kappa (1+w)T^{\frac{1}{w}}\,\, and \,\, \varepsilon = \kappa T^{\frac{(1+w)}{w}}
\label{sol-negativa}
\end{equation}
where $\kappa$ is a constant.
This branch of solutions was adopted, among others, by \cite{lima04} and \cite{german06}
in their analyses of the entropy evolution in a world model dominated by a phantom field.
Notice that since the energy density as a function of the
scale factor varies as $\varepsilon \propto a^{-3(1+w)}$, the entropy per comoving
unit volume is constant, e.g., $sa^3=constant$. In reference \cite{german06}
the variation of the total entropy, was defined by
the sum of the phantom fluid entropy inside the event horizon plus the
area entropy of the
event horizon itself as suggested by \cite{davies87}. This approach requires an additional
hypothesis concerning the temperature characterizing the fluid, usually assumed to be
equal the horizon temperature given by the Gibbons-Hawking relation (\cite{gibbons77}).
Under these conditions, the total entropy is always zero and the GSL is
verified \cite{german06},
including the de Sitter case \cite{pollock89}. The assumption of thermal equilibrium with
the horizon requires that $\mid R_H/\dot R_H\mid > (R_H/c)$ or, in other words,
the time scale
in which the event horizon varies should be larger than the radiation crossing time, a
condition satisfied only for cosmologies with small departures from the de Sitter model.
Moreover, if we require further that the characteristic wavelength of the field quanta be
smaller than the horizon radius, then necessarily the fluid temperature must be higher
than the Gibbons-Hawking temperature.

 The second solution is expressed by the pair \cite{gon04}
\begin{equation}
s = \kappa\left[\frac{T}{(1+w)}\right]^{\frac{1}{w}}\,\, and \,\,
\varepsilon = \kappa\left[\frac{T}{(1+w)}\right]^{\frac{(1+w)}{w}}
\label{sol-positiva}
\end{equation}
where now the temperature is negative. In spite of not being
common in physics, this concept is not meaningless. Experiments on
the nuclear spin relaxation of a LiF crystal, after exposition in
a magnetic field, indicate that the spin state is properly
described by a negative spin temperature, since the system loses
internal energy as it gains entropy \cite{pur51}. Different
authors have addressed to the thermodynamic properties of systems
at negative temperatures (see, for instance,\cite{ram56}). The
basic requirement for the existence of a negative temperature is
that the entropy density should not be a monotonically increasing
function of the energy density. This is exactly the situation
resulting from eqs.~\ref{sol-positiva}, since $ds/d\varepsilon
\propto \varepsilon^{-\frac{w}{(1+w)}}/(1+w) < 0$ for a phantom
field and such a slope is proportional to $T^{-1}$. It is worth
mentioning that negative temperatures are not colder than absolute
zero but instead are hotter than infinite temperatures. As a
consequence, if a system with negative temperature interacts with
another system having a positive temperature, the energy will
always flow from the former to the latter.

\subsection{Accreting a phantom field}

In the light of the above considerations, a BH (having $T>0$, and therefore acting as
a source) embedded in a fluid with a negative
temperature will absorb energy. The problem of a scalar field in the
presence of a BH was already considered
by different authors. References \cite{jacobson99} and \cite{bean02} discussed
fields with positive kinetic energy by
solving the wave equation in a Schwarzschild background and concluding that
the accretion rate
depends only on the kinetic term (see also \cite{horvath} and \cite{harada}). A
different approach was
adopted by \cite{babi04}, who considered
stationary solutions for the accretion of a fluid onto a
Schwarzschild BH, generalizing the early results
by \cite{michel72} and concluding that the accretion rate is proportional to the quantity
$(P+\varepsilon)$, which is equal to twice the kinetic
energy of a homogeneous scalar field and negative
for a phantom field. These results are a simple consequence of
the first law: the amount of energy $dE$ crossing the BH horizon
in a time interval $dt$ is
\begin{equation}
dE = 4\pi r_g^2(P+\varepsilon)cdt
\end{equation}
where $r_g=2GM_{bh}/c^2$ is the gravitational radius. The absorbed energy
produces a variation of the BH entropy equal to
\begin{equation}
dS_{bh}=\frac{8\pi GM}{\hbar c}\frac{dM}{dt}dt
\end{equation}
Since $T = dE/dS$, using the eqs. above and the well known relation
for the BH temperature, e.g.,
$kT = \hbar c^3/(8\pi GM)$, one obtains
\begin{equation}
\frac{dM}{dt}= \frac{16\pi (GM_{bh})^2}{c^5}\dot\phi^2
\end{equation}
which coincides with rates derived by other methods mentioned above
and gives consistency to the thermodynamic approach.

Consider now a comoving volume $V \propto a^3$ containing
a phantom fluid and a black hole of mass $M_{bh}$ (the generalization for
the case including $N$ black holes
is immediate and trivial). Assuming further that distortions in the
spacetime inside the cavity, due to the
presence of the black hole, can be neglected in a first approximation, the total
entropy can be written as
\begin{equation}
S=\frac{4\pi GM^2_{bh}}{\hbar c}+ \kappa\varepsilon^{\frac{1}{(1+w)}}V
\label{entropy1}
\end{equation}
where the first term represents the black hole
entropy and the second, the phantom fluid entropy inside the
comoving volume V. Due to the accretion process, in a short time interval
the BH mass varies by an amount $\Delta M_{bh}$ and the energy
density of the phantom field varies by an
amount $\Delta\varepsilon$. Under these conditions, using
eq.~\ref{entropy1}, the total entropy variation in the cavity is
\begin{equation}
\Delta S=\frac{8\pi GM_{bh}}{\hbar c}\Delta M_{bh}+\frac{\kappa}{(1+w)}
\varepsilon^{-\frac{w}{(1+w)}}\Delta\varepsilon V
\label{entropy2}
\end{equation}
For a phantom fluid modeled by a scalar field, only the kinetic term
contributes to the accretion
as discussed above, thus, energy conservation inside the cavity implies that
\begin{equation}
c^2\Delta M_{bh}=-\frac{1}{2}\Delta\dot\phi^2V=-\frac{1}{2}(1+w)\Delta\varepsilon V
\label{deltas}
\end{equation}
The above equation says that if $w>-1$ (quintessence fluid) a negative variation of
the field energy density
implies a positive variation in the black hole mass whereas if $w<-1$ (phantom field), a
decrease in the field energy density implies also a decrease in the black
hole mass since the kinetic term is now negative.
Using the above result, eq.~\ref{entropy2} can be recast as
\begin{equation}
\Delta S=\left[\frac{8\pi GM_{bh}}{\hbar c}-\frac{2\kappa c^2}{(1+w)^2}
\varepsilon^{-\frac{w}{(1+w)}}\right]\Delta M_{bh}
\label{entropy3}
\end{equation}
Since in the accretion process $\Delta M_{bh} < 0$, in order to satisfy
the GSL it is required that
\begin{equation}
M_{bh,crit} \leq \frac{\kappa\hbar c^3}{4\pi G(1+w)^2}\varepsilon^{-\frac{w}{(1+w)}}
\label{critical1}
\end{equation}
This relation implies that there is a critical mass above which the BH cannot
accrete the phantom fluid,
otherwise the GSL is violated. The existence of a critical mass can be easily
understood. As the BH
accretes phantom energy its entropy decreases. Thus, in order to have an
increase in the total entropy, the
field entropy must increase and compensate the loss in the BH entropy. This is
possible only if the
phantom fluid has a negative temperature, since in this case a decrease in
the energy density increases
the entropy. However, negative variations in the BH entropy are
proportional to the BH mass and, above a certain
value, they cannot be counterbalanced by increasing the field entropy. Such a
situation would be completely
different had we adopted the solution in which the phantom
fluid has a positive temperature. In this case, using the
same reasoning as before leads to the condition for the validity of the GSL
\begin{equation}
\frac{8\pi GM_{bh}}{\hbar c}-\frac{2\kappa c^2}{(1+w)}
\varepsilon^{-\frac{w}{(1+w)}} \leq 0
\end{equation}
This condition cannot be satisfied since the second term on
the left side of the above equation is
always positive when $w<-1$. Thus, the negative entropy solution for
the phantom fluid has more drastic
consequences since it implies that the accretion process is not possible
unless the GSL be violated.

A more quantitative analysis of the fate of BHs embedded in a
phantom field requires the knowledge of
its entropy, which remains indeterminate since the constant $\kappa$ is unknown.
However, an upper bound can be derived from the holographic principle
which states that
\begin{equation}
\frac{4\pi}{3}R^3\kappa\varepsilon^{\frac{1}{(1+w)}} \leq \frac{\pi}{l^2_P}R^2
\label{bound1}
\end{equation}
where $R$ is the event horizon radius
and $l_P$ is
the Planck's length. Using the time solutions for the energy
density and the event horizon radius
in a phantom dominated universe (see, for instance, \cite{gon04}), eq.~\ref{bound1}
can be recast as
\begin{equation}
\left[\frac{(t_*-t_0)}{(t_*-t)}\right]^{\frac{(1-w)}{(1+w)}} \leq
\frac{3\mid 1+3w\mid}{8}\frac{\sqrt{\Omega_{\Lambda}}H_0}{cl_P^2s_0}
\label{bound2}
\end{equation}
where $t_*$ is the big rip time, $t_0$ is the present age of the universe and
$s_0 =\kappa\varepsilon_0^{\frac{1}{(1+w)}}$ is the present phantom entropy
density. If $w<-1$,
the left side of eq.~\ref{bound2} goes to zero as the universe expands and approaches the
singularity, whereas it is equal to the unity at the present time. Thus, the
holographic bound is satisfied if
\begin{equation}
s_0 \leq \frac{3\mid 1+3w\mid}{8}\frac{\sqrt{\Omega_{\Lambda}}H_0}{cl_P^2}
\label{bound3}
\end{equation}
or, numerically
\begin{equation}
s_0 \simeq 9.37\times 10^{36}\mid 1+3w\mid\, cm^{-3}
\end{equation}
Using entropy conservation within a comoving volume and the entropy bound
given by eq.~\ref{bound3}
into eq.~\ref{critical1} a more generous upper limit can be derived, e.g.,
\begin{equation}
M_{bh,crit} \leq \frac{2GM_P^2}{3(1+w)^2}\frac{s_0}{\Omega_{\Lambda}H^2_0}
\left[\frac{(t_*-t)}{(t_*-t_0)}\right]^{\frac{2w}{(1+w)}}
\label{critical2}
\end{equation}
where $M_P$ is the Planck's mass.

In order to perform some numerical estimates, let us suppose $w=-3/2$, the same value
adopted by \cite{cal03} in their calculations, implying $(t_*-t_0) \simeq 22.7~Gyr$.
Presently, the critical mass is about $3.6\times 10^{23}\, M_{\odot}$, a quite huge
value, allowing all BHs in the universe to accrete the phantom field. However, for
the adopted value
of the equation of state parameter, the critical mass decreases as $(t_*-t)^6$ and
is about $10^9 M_{\odot}$ at 85.2 Myr before the big rip. Thus, a BH having presently
such a mass, will still accrete the phantom field beyond that time only if its mass
has substantially decreased, otherwise the limit imposed by the GSL avoids further
accretion. The mass evolution of the BH due to accretion of the phantom fluid is given
by

\begin{equation}
M_{bh}(t)=\frac{M_{bh,0}}{ 1+\frac{8 t_P}{3|1+w|}
\frac{M_{bh,0}}{M_P} \times
{\left[\frac{1}{(t_{*}-t)}-\frac{1}{(t_{*}-t_{0})} \right]}}
\end{equation}

where $t_P$ is the Planck time scale. The above equation shows
that only near the singularity the mass of the BH is substantially
altered and that the evolution in these late phases is independent
of the initial BH mass $M_{bh,0}$. A simple calculation indicates
that a BH with present mass of $10^9 M_{\odot}$ will suffer a
relative mass loss, up to 85.2 Myr before the singularity, of
about $\Delta M_{bh}/M_{bh,0} \simeq 10^{-11}$, a quite
insignificant amount. This means that, just after reaching the
critical mass limit, the accretion process stops. Estimates for
lower mass BHs lead to similar results.

\section{Conclusions}

For phantom fields, the violation of the dominant energy condition leads to two
alternatives: either the entropy density is negative, being the temperature
positive or the entropy density is positive but in this case the associated
temperature is negative. In the latter case, if the GSL is supposed to be
valid, then there is a critical BH mass above which the accretion process is
not allowed. However, the accretion process is significant only
near the singularity and, as
a consequence BHs reach the critical mass value before the big rip, having lost a
negligible amount of mass. If the negative entropy solution is adopted, the GSL forbids
the accretion process. In both cases, only a violation of the GSL is consistent with
the scenario devised by \cite{babi04}.


\section*{References}

\begin{thebibliography}{25}
\bibitem{alam06}
Alam, U., Sahi, V. and Starobinsky, A.A., 2007, {\it JCAP} {\bf 0702:011}, (astro-ph/061238)
\bibitem{babi04}
Babichev, E., Dokuchaev, V. and Eroshenko, Yu., 2004, {\it Phys.Rev.Lett.} {\bf 93}, 021102
\bibitem{babi05}
Babichev, E., Dokuchaev, V. and Eroshenko, Yu., 2005, {\it J.Exp.Theor.Phys.} {\bf 100} 528
\bibitem{bean02}
Bean, R. and Magueijo, J., 2002, {\it Phys.Rev.} {\bf D66}, 063505
\bibitem{bek74}
Bekenstein, J.D., 1974, {\it Phys.Rev.} {\bf D9}, 3292
\bibitem{bek81}
Bekenstein, J.D., 1981, {\it Phys.Rev.} {\bf D23}, 287
\bibitem{cal03}
Caldwell, R.R., Kamionkowsky, M. and Weinberg, N.N., 2003, {\it Phys.Rev.Lett.} {\bf 91}, 071301
\bibitem{cline04}
Cline, J.M., Jeon, S. and Moore, G.D., 2004, {\it Phys.Rev.} {\bf D70}, 043543
\bibitem{horvath}
Cust\'odio, P.S. and Horvath, J.E., 2005, {\it Int.Jour.Mod.PHys.} {\bf D14}, 257
\bibitem{davies87}
Davies, P.C.W., 1987, {\it Class.Quantum Grav.} {\bf 4}, L225
\bibitem{deu02}
Deustua, S.E., Caldwell, R., Garnavich, P., Hui, L. and Refregier, A., 2002,
{\it Preprint} astro-ph/0207293
\bibitem{fram03}
Frampton, P., 2003, {\it Phys.Lett.} {\bf B545}, 139
\bibitem{gibbons77}
Gibbons, G. and Hawking, S.W., 1977, {\it Phys.Rev.} {\bf D15}, 2738
\bibitem{gon04}
Gonz\'alez-Diaz, P.F. and Siguenza, C.L., 2004, {\it Nucl.Phys.} {\bf B697}, 363
\bibitem{harada}
Harada, T. and Carr, B.J., 2005, {\it Phys.Rev.} {\bf D71}, 104010
\bibitem{har95}
Harrison, E.R., 1995, {\it Astrophys. J.} {\bf 446}, 63
\bibitem{german06}
Izquierdo, G. and Pav\'on, D., 2006, {\it Phys.Lett.} {\bf B633}, 420
\bibitem{jacobson99}
Jacobson, T., 1999, {\it Phys.Rev.Lett.} {\bf 83}, 2699
\bibitem{lima04}
Lima, J.A.S and Alcaniz, J.S., 2004, {\it Phys.Lett.} {\bf B600}, 191
\bibitem{michel72}
Michel, F.C., 1972, {\it Astroph.Sp.Sci.} {\bf 15}, 153
\bibitem{nil84}
Nilles, H.P., 1984, {\it Phys.Rep.} {\bf 110}, 1
\bibitem{nojiri04}
Nojiri, S. and Odintsov S.D., 2004, {\it Phys.Rev.} {\bf D70}, 103522
\bibitem{perl99}
Perlmutter, S. et al., 1999, {\it Astrophys. J.} {\bf 517}, 565
\bibitem{pollock89}
Pollock, M.D. and Singh, T.P., 1989, {\it Class. Quantum Grav.} {\bf 6}, 901
\bibitem{pur51}
Purcell, E.M. and Pound, R.V., 1951, {\it Phys.Rev.} {\bf 81}, 279
\bibitem{ram56}
Ramsey, N.F., 1956, {\it Phys.Rev.} {\bf 103}, 20
\bibitem{riess98}
Riess, A.G. et al., 1998, {\it Astron. J.} {\bf 116}, 1009
\bibitem{tonry03}
Tonry, J.L. et al., 2003, {\it Astrophys. J.} {\bf 594}, 1


\end{thebibliography}
\end{document}